\documentclass[twocolumn,showpacs,superscriptaddress,nofootinbib,prd]{revtex4}

\usepackage{graphicx}
\newcommand{\ra}{\rightarrow}
\newcommand{\conv}{\mbox{\scriptsize$\otimes$}}

\begin{document}
\title{Effect of polarized gluon distribution on spin asymmetries \\
for neutral and charged pion production}
\author{M. Hirai}
\email[E-mail: ]{mhirai@rarfaxp.riken.jp}
\affiliation{Institute of Particle and Nuclear Studies, 
          High Energy Accelerator Research Organization, \\
          1-1, Ooho, Tsukuba, Ibaraki, 305-0801, JAPAN}
\author{K. Sudoh}
\email[E-mail: ]{sudou@rarfaxp.riken.jp}
\affiliation{Radiation Laboratory, 
RIKEN (The Institute of Physical and Chemical Research), \\
Wako, Saitama 351-0198, JAPAN}
\date{\today}

\begin{abstract}
A longitudinal double spin asymmetry for $\pi^0$ production has been 
measured by the PHENIX collaboration.
The asymmetry is sensitive to the polarized gluon distribution 
and is indicated to be positive by theoretical predictions.
We study a correlation between behavior of the asymmetry and polarized 
gluon distribution in neutral and charged pion production at RHIC. 
\end{abstract}

\maketitle

\section{Introduction}
Determination of the polarized parton distribution functions (PDFs) is 
crucial for understanding the spin structure of the nucleon \cite{SPIN}.
As is known well, the proton spin is composed of the spin 
and orbital angular momentum of quarks and gluons.
Several parametrizations of the polarized PDFs have been proposed, 
and have successfully reproduced experimental data \cite{AAC03,BB,LSS,GRSV,DS}.
In particular, the amount of the proton spin carried by quarks is determined well 
by global analyses with the polarized deep inelastic scattering (DIS) data. 
The value is about $\Delta\Sigma=0.1\sim0.3$, whereas the prediction from 
the naive quark model is $\Delta\Sigma =1$.
This surprising result leads to extensive study on the gluon polarization.
The current parametrizations suggest a large positive polarization of gluon.
However, 
our knowledge about the polarized gluon distribution $\Delta g(x,Q^2)$ is still poor, 
since theoretical and experimental uncertainties are rather large.
The determination of $\Delta g(x,Q^2)$ gives us a clue to the proton spin puzzle.

The RHIC is the first high energy polarized proton-proton collider 
to measure $\Delta g(x,Q^2)$ \cite{RHIC}.
We can extract information about $\Delta g(x,Q^2)$ through various processes, 
e.g., prompt photon production, jet production, and heavy flavor production. 
These processes are quite sensitive to $\Delta g(x,Q^2)$, 
since gluons in the initial state associate with the cross section 
in leading order (LO).

Recently, the PHENIX collaboration has reported results 
for inclusive $\pi^0$ production $pp\ra \pi^0 X$ \cite{PHENIX03_pol} 
which is also likely to be sensitive to $\Delta g(x,Q^2)$. 
The double spin asymmetry was measured in longitudinally polarized 
proton-proton collisions at RHIC in the kinematical ranges: center-of-mass 
(c.m.) energy $\sqrt{s}=200$ GeV and central rapidity $|\eta|\leq 0.38$.
The data imply that the asymmetry might be negative 
at transverse momentum $p_T = 1\sim3$ GeV.
The lower bound of the $\pi^0$ asymmetry at low $p_T$ has been considered, 
and a slight negative asymmetry by modifying $\Delta g(x,Q^2)$ 
has been demonstrated in Ref. \cite{JSKW}.
However, there is no theoretical predictions indicating 
large negative asymmetry.

In this paper, we study the behavior of the $\pi^0$ double spin asymmetry 
correlated with $\Delta g(x,Q^2)$ in Sec \ref{asymmetry}.
By using three types of $\Delta g(x,Q^2)$, 
we suggest that the asymmetry in large $p_T$ region is 
more sensitive to the functional form of $\Delta g(x,Q^2)$. 
An impact of the new data on determination of $\Delta g(x)$ is discussed
in terms of uncertainty of the asymmetry coming from the polarized PDFs.
Furthermore, we discuss a spin asymmetry for charged pion production 
in Sec. \ref{chargedAsym}. 
An asymmetry taking the difference of cross sections for $\pi^+$ and $\pi^-$ 
production is proposed, 
and it is useful to discuss the sign of $\Delta g(x)$ in the whole $p_T$ region.
The Summary is given in Sec. \ref{Summary}.

\section{\label{asymmetry} Spin asymmetry for neutral pion production}
\subsection{Ambiguity of the polarized cross section}
First, we describe the longitudinal double spin asymmetry for $\pi^0$ production. 
It is defined by 
\begin{equation}
A_{LL}^{\pi^0}\equiv
\frac{[d\sigma_{++}-d\sigma_{+-}]/dp_T}{[d\sigma_{++}+d\sigma_{+-}]/dp_T}
=\frac{d\Delta\sigma/dp_T}{d\sigma/dp_T} ,
\end{equation}
where $p_T$ is the transverse momentum of produced pion.
$d\sigma_{hh'}$ denotes the spin-dependent cross section with definite 
helicity $h$ and $h'$ for incident protons.

The cross sections can be separated short distance parts 
from long distance parts by the QCD factorization theorem.
The short distance parts represent interaction amplitudes of hard partons, 
and are calculable in the framework of perturbative QCD (pQCD).
On the other hand, the long distance parts such as PDFs and fragmentation functions 
should be determined by experimental data.
The polarized cross section $\Delta\sigma$ is written by
\begin{eqnarray}
\frac{d\Delta\sigma^{pp\ra \pi^0 X}}{dp_T}
&=&\sum_{a,b,c}\int^{\eta^{\max}}_{\eta^{\min}}d\eta
\int^{1}_{x_a^{\min}}dx_a \int^{1}_{x_b^{\min}}dx_b \nonumber \\
&&\times\Delta f_{a}(x_a,Q^2) \Delta f_{b}(x_b,Q^2) \nonumber \\
&&\times
{\cal J}\left(\frac{\partial(\hat{t},z)}{\partial(p_T,\eta)}\right)
\frac{\Delta\hat{\sigma}^{ab\ra cX}(\hat{s},\hat{t})}{d\hat{t}} 
\nonumber \\ 
&&\times D_{c}^{\pi^0}(z,Q^2), 
\label{eq:xsec}
\end{eqnarray}
where $\Delta f_i(x,Q^2)$ is the polarized PDFs, and 
$D_{c}^{\pi^0}(z,Q^2)$ is the spin-independent fragmentation function 
decaying into pion $c\ra\pi^0$ with a momentum fraction $z$.
The sum is over the partonic processes $a+b\ra c+X$ associated with 
$\pi^0$ production. 
${\cal J}$ is the Jacobian which transforms kinematical variables from 
$\hat{t}$ and $z$ into $p_T$ and $\eta$ of the produced $\pi^0$.
$\Delta\hat{\sigma}$ describes the polarized cross section of subprocesses. 
The partonic Mandelstam variables $\hat{s}$ and $\hat{t}$ are defined by 
$\hat{s}=(p_a + p_b)^2$ and $\hat{t}=(p_a - p_c)^2$ with partonic momentum 
$p_i$, respectively.
The squared c.m. energy $s$ is related to $\hat{s}$ through $\hat{s}=x_a x_b s$ 
and set as $\sqrt{s}=200$ GeV.
The pseudo-rapidity $\eta$ is limited as $|\eta|\leq0.38$ in the PHENIX acceptance.

In this analysis, 
the cross sections and the spin asymmetry are calculated in LO level.
Rigorous analysis of ${\cal O}(\alpha_s^3)$ next-to-leading order (NLO) 
calculation has been established in Ref. \cite{JSSW}.
We believe that the qualitative behavior of the asymmetry does not change, 
even if NLO corrections are included in our study.
In numerical calculations, we adopt the AAC set \cite{AAC03} as the 
polarized PDFs and the Kretzer set \cite{Kretzer00} as the fragmentation functions.
We choose the scale $Q^2=p_T^2$.

The partonic subprocesses in LO are composed of 
${\cal O}(\alpha_s^2)$ $2\ra 2$ tree-level channels listed as 
$gg\ra q(g)X$, $qg\ra q(g)X$, $qq\ra qX$, 
$q\bar{q}\ra q(g, q')X$, $qq'\ra qX$, and $q\bar{q}'\ra qX$ including 
channels of the permutation $q\leftrightarrow \bar{q}$.
Main contribution to the polarized cross section comes from $gg\ra q(g)X$ 
and $qg\ra q(g)X$ channels with conventional PDFs and fragmentation functions. 
The $gg$ contribution dominates in low $p_T$ region 
and steeply decreases with $p_T$ increases.
Then, the $qg$ process becomes dominant in larger $p_T$ region.
The crossing point of these contributions however depends on parametrization 
of the polarized PDFs.
In both cases, the spin asymmetry for $\pi^0$ production is sensitive to 
the gluon polarization.

As mentioned above, the partonic cross section $\Delta\hat{\sigma}$ is 
well-defined in the pQCD framework.
Hence, as a cause of inconsistency with the PHENIX data, 
we consider the ambiguity of long distance parts: 
fragmentation functions and PDFs.

The fragmentation into $\pi^0$ includes all channels $q,\bar{q},g\ra \pi^0$.
Each component of the fragmentation functions $D_{c}^{\pi^0}$ can be determined 
by global analyses with several experiments \cite{Kretzer00,KKP00}.
The unpolarized cross section measured by the PHENIX 
\cite{Phenix03_unpol} are consistent with NLO pQCD calculations 
within model dependence of $D_{c}^{\pi^0}$. 
These precise measurements give strong constraint on the fragmentation functions. 
Significant modification of them would not be expected. 
Therefore, the fragmentation functions are not the source of the negative asymmetry 
even if they have uncertainty to some extent. 

In the polarized reaction, kinematical ranges and 
the fragmentation functions are the same as the unpolarized case 
except the polarized PDFs.
For the polarized quark distributions $\Delta q(x)$ and $\Delta \bar{q}(x)$, 
the antiquark distributions and their flavor structure are not well known.
For $\pi^0$ production, subprocesses are (light quark) flavor blind 
reaction, and the predominant $qg$ process depends on the sum 
$\Delta q(x)+\Delta \bar{q}(x)$ which is relatively determined well 
by the polarized DIS data \cite{JSKW}, 
and so ambiguities of the polarized quark distributions can be neglected. 
In consequence, the undetermined polarized gluon distribution $\Delta g(x)$
remains as the source of the uncertainty of the asymmetry.

\subsection{Correlation between the spin asymmetry 
and the polarized gluon distribution $\Delta g(x)$}
\begin{figure}[b]
\includegraphics[scale=0.42,clip]{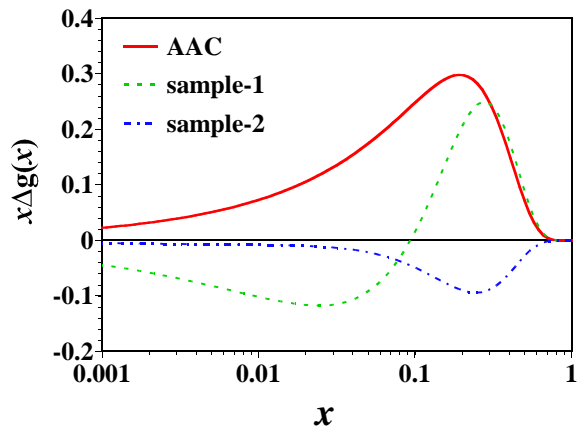}
\caption{\label{fig-gluon}
Polarized gluon distributions $\Delta g(x)$ at $p_T =2.5$ GeV.
Solid, dashed, and dot-dashed curves indicate 
the AAC, sample-1, and 2 distributions, respectively.}
\end{figure}

To investigate a role of $\Delta g(x)$ for the behavior of the asymmetry, 
we prepare three functional forms as shown in Fig \ref{fig-gluon}.
Solid curve shows $\Delta g(x)$ by the global analysis 
with the polarized DIS data \cite{AAC03}.
Dashed and dot-dashed curves show two artificial modified $\Delta g(x)$, 
respectively.
The sample-1 distribution has a node. 
The gluon distribution with a node has been indicated in the paper 
by J\"ager {\it et al.,} \cite{JSKW}. 
Our distribution is negative in the small $x$ region, 
and positive in the large $x$ region. 
It has opposite signs of $\Delta g(x)$ shown in Fig. 2 of their paper. 
The sample-2 distribution is small negative in the whole $x$ region. 
Their distribution is similar to the sample-2 rather than the sample-1. 
It shows barely positive at small $x$, while the sample-2 is negative.
%
Since the sample-1 and 2 are within the $\Delta g(x)$ uncertainty 
by the AAC analysis, 
these distributions can be adopted as a model of $\Delta g(x)$.
These are taken account of the $Q^2$ dependence 
by the Dokshitzer-Gribov-Lipatov-Altarelli-Parisi (DGLAP) equation 
with the polarized quark and antiquark distributions.

\begin{figure}[t]
\includegraphics[scale=0.42,clip]{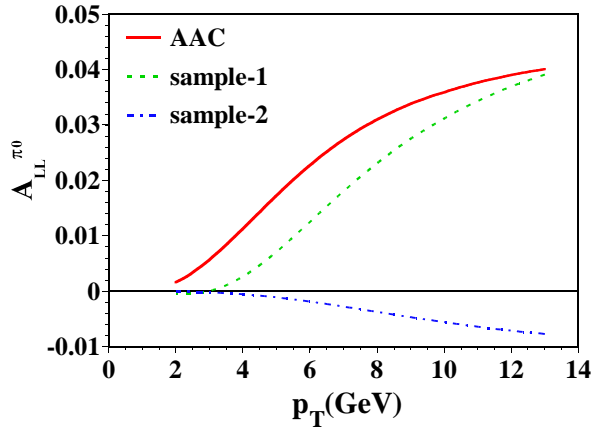}
\caption{Spin asymmetries for $\pi^0$ production 
by using three different $\Delta g(x)$ in Fig. \ref{fig-gluon}.
}
\label{fig-All}
\end{figure}

We discuss the behavior of the spin asymmetry 
associated with the functional form of $\Delta g(x)$. 
The obtained asymmetries with these gluon distributions are shown 
in Fig. \ref{fig-All}.
We find that 
the asymmetry for the AAC $\Delta g(x)$ is positive in the whole $p_T$ region. 
The asymmetries for the sample-1 and 2 become negative at low $p_T$. 
In particular, we obtained the negative asymmetry in the whole $p_T$ region 
by using the sample-2 $\Delta g(x)$.
Furthermore, one can see variations of these asymmetries at large $p_T$.

The asymmetry for the AAC is positive and increases with $p_T$.
The positive polarization for $\Delta g(x)$ generates 
positive contributions of $gg$ and $qg$ processes 
which dominate in the $\pi^0$ production.
In this case, the asymmetry cannot become negative.

The positive $\Delta g(x)$ is suggested by 
the recent global analyses with the polarized DIS data \cite{AAC03, BB, LSS, GRSV, DS}.
Although these analyses obtain good agreement with the experimental data, 
the $\Delta g(x)$ cannot be determined and it has large uncertainty. 
Therefore, we cannot rule out the negative polarization for $\Delta g(x)$. 
There is a possibility of the negative asymmetry 
with the modified $\Delta g(x)$. 

For the sample-1 in Fig. \ref{fig-All}, 
the asymmetry is slight negative at low $p_T$ 
and changes into positive at $p_T=3$ GeV.
As is mentioned in Ref. \cite{JSKW}, 
the $\Delta g(x)$ with a node has the possibility of making
the small negative asymmetry at low $p_T$.
In the region $p_T<3$ GeV, 
we find that contributions of $gg$ and $qg$ processes are negative, respectively.
To make negative $gg$ contribution 
would be needed opposite polarizations of $\Delta g(x)$ at $x_a$ and $x_b$.
Computed by using several shapes of $\Delta g(x)$ with a node, 
the $gg$ contribution is not always negative.
The contribution basically depends on the shape of $\Delta g(x)$ 
even if it has a node.

In the region $p_T>3$ GeV, 
the $gg$ contribution changes into positive, 
and dominates in the region $p_T <10$ GeV. 
This is because that the node rapidly shifts toward low-$x$ direction 
due to $Q^2$ evolution with $p_T$.
Therefore, the positive polarization for $\Delta g(x)$ at medium $x$ 
contributes predominantly to the positive asymmetry via the $gg$ process.
Furthermore, the asymmetry at large $p_T$ is sensitive to 
the behavior of $\Delta g(x)$ at medium $x$.

As another possibility of the negative asymmetry, 
we choose slight negative polarization for $\Delta g(x)$.
In this case, the $gg$ contribution is positive 
while the $qg$ contribution is negative.
The asymmetry is determined by the difference between two contributions.  
The $gg$ and $qg$ contributions are proportional to $(\Delta g)^2$ 
and $\Delta g$, respectively. 
The $gg$ contribution is more sensitive to the behavior of $\Delta g(x)$.
In particular, 
the behavior at low $x$ significantly affects on the contribution at low $p_T$ 
since the value of $x^{\min}$ in Eq. (\ref{eq:xsec}) is rather small.
In order to make the positive $gg$ contribution smaller,
the $\Delta g(x)$ for the sample-2 is taken small polarization at low $x$ 
as shown in Fig. \ref{fig-gluon}.

In Fig. \ref{fig-All}, as far as the sample-2 is concerned, 
the asymmetry indeed becomes negative in the whole $p_T$ region.
In the region $p_T <3$ GeV, 
the small negative polarization for $\Delta g(x)$ 
generates slight positive contribution of the $gg$ process.
In this case, 
the $gg$ contribution is the same order of magnitude as the $qg$ contribution, 
and almost cancel out the negative contribution.
The asymmetry is therefore determined by other processes.
The total contribution of the processes except above two processes 
becomes slight negative.
Above the region,
the $gg$ contribution rapidly decreases with $p_T$ increases.
The $qg$ process becomes dominant contribution, 
which provides the negative asymmetry \cite{JSKW}.
Thus, the negative asymmetry can be obtained in the whole $p_T$ region 
by using the negative $\Delta g(x)$ 
which makes the $qg$ contribution larger than the $gg$ contribution.

In the sample-2, we should note that
the magnitude of $\Delta g(x)$ at the minimum point cannot be large. 
This is because that the shape of $\Delta g(x)$ is rapidly varied 
by the $Q^2$ evolution,  
the minimum point of $\Delta g(x)$ shifts toward low-$x$ 
and the width broadens.
At moderate $p_T$, 
the $gg$ process is more sensitive to 
the low-$x$ behavior of the evolved $\Delta g(x)$ than the $qg$ process.
If the $\Delta g(x)$ is taken large negative polarization at the minimum point, 
the magnitude of the $gg$ contribution becomes rapidly large 
compared with the $qg$ contribution, 
and then the asymmetry becomes positive at moderate $p_T$.
The small negative $\Delta g(x)$ is therefore required 
to obtain the negative asymmetry in the whole $p_T$ region. 

In above two cases at low $p_T$, 
we cannot also obtain negative value exceeded the lower limit $-0.1\%$ 
that is suggested in Ref. \cite{JSKW}.
Furthermore, even if the asymmetry is positive, 
the magnitude is below $1 \%$.
As we discussed, 
the functional form of $\Delta g(x)$ needs some restraints 
to make the asymmetry negative. 
It is difficult to obtain sizable value 
in comparison with the positive case.

At large $p_T$, 
the difference of the obtained asymmetries remarkably reflects 
the medium-$x$ behavior of $\Delta g(x)$.
Experimental data in the region is useful to determine the $\Delta g(x)$.
For instance, the asymmetry for the sample-2 becomes rather larger 
to negative direction.
If future precise data indicate the negative asymmetry in the region, 
the $\Delta g(x)$ requires significant modification of its functional form.
It has the potential of the negative gluon contribution to the nucleon spin.
In order to understand the behavior of $\Delta g(x)$ in detail, 
we require experimental data covering a wide $p_T$ region.

\subsection{Uncertainty of the spin asymmetry}
Next, we consider the effect of the $\pi^0$ data
on the $\Delta g(x)$ determination 
in terms of the uncertainty estimation for the spin asymmetry.
The large uncertainty of $\Delta g(x)$ 
implies the difficulty of extracting the gluon contribution 
from the polarized DIS data.
We are therefore interested in constraint power of the new data on $\Delta g(x)$. 
If the experimental data are included in a global analysis, 
the asymmetry uncertainty will be bounded within statistical error range. 
See, for example, Fig. 2 of Ref. [2]. 
As far as evaluation of the constraint is concerned, 
the uncertainty can be compared with statistical errors of the data, 
although it is rough evaluation.  

The asymmetry uncertainty coming from the polarized PDFs is defined 
by a polarized cross section uncertainty 
divided by a unpolarized cross section: 
$\delta A_{LL}^{\pi^0}=\delta \Delta \sigma^{\pi^0}/\sigma^{\pi^0}$.
The cross section uncertainty is obtained by taking 
a root sum square of uncertainties of all subprocesses. 
These uncertainties are estimated by the Hessian method, 
and are given by
\begin{equation}
  \left[\delta \Delta \sigma_k^{\pi^0} \right]^2=
       \Delta \chi^2 \sum_{i,j}
       \left( \frac{\partial \Delta \sigma_k^{\pi^0}(p_{_T})}
                   {\partial a_i} \right)
       H_{ij}^{-1}
       \left( \frac{\partial \Delta \sigma_k^{\pi^0}(p_{_T})}
                   {\partial a_j} \right) \ ,
    \label{eq:erroe-M}
\end{equation}
where $k$ is the index of subprocesses. 
$a_i$ is a optimized parameter in the polarized PDFs.
$H_{ij}$ is the Hessian matrix 
which has the information of the parameter errors and the correlation 
between these parameters.
The $\Delta \chi^2$ determines a confidence level of the uncertainty, 
and is estimated so that the level corresponds to the 1$\sigma$ standard error.
We choose the same value as the AAC analysis \cite{AAC03}. 
Further, the gradient terms for the subprocesses
$\partial \Delta \sigma_k^{\pi^0}(p_{_T})/\partial a_i$ 
are obtained by
\begin{eqnarray}
\frac{d\Delta\sigma_k^{\pi^0}}{dp_T}
	&&=\sum_{a,b,c}\int^{\eta^{\max}}_{\eta^{\min}}d\eta
	    \int^{1}_{x_a^{\min}}dx_a \int^{1}_{x_b^{\min}}dx_b \nonumber \\
	&& \times \left[ 
              \frac{\partial \Delta f_a(x_a)}{\partial a_i} \Delta f_b(x_b)
              +\Delta f_a(x_a)\frac{\partial \Delta f_b(x_b)}{\partial a_i} 
        \right] \nonumber \\
	&&\times
        {\cal J}\left(\frac{\partial(\hat{t},z)}{\partial(p_T,\eta)}\right)
	  \frac{\Delta\hat{\sigma}^{ab\ra cX}(\hat{s},\hat{t})}{d\hat{t}} 
	D_{c}^{\pi^0}(z), 
\end{eqnarray}
The gradient terms for the polarized PDF $\partial \Delta f_a(x_a)/\partial a_i$ 
are analytically obtained at initial scale $Q_0^2$, 
and are numerically evolved to arbitrary scale $Q^2$ 
by the DGLAP equation. 

\begin{figure}[t]
        \includegraphics[scale=0.42,clip]{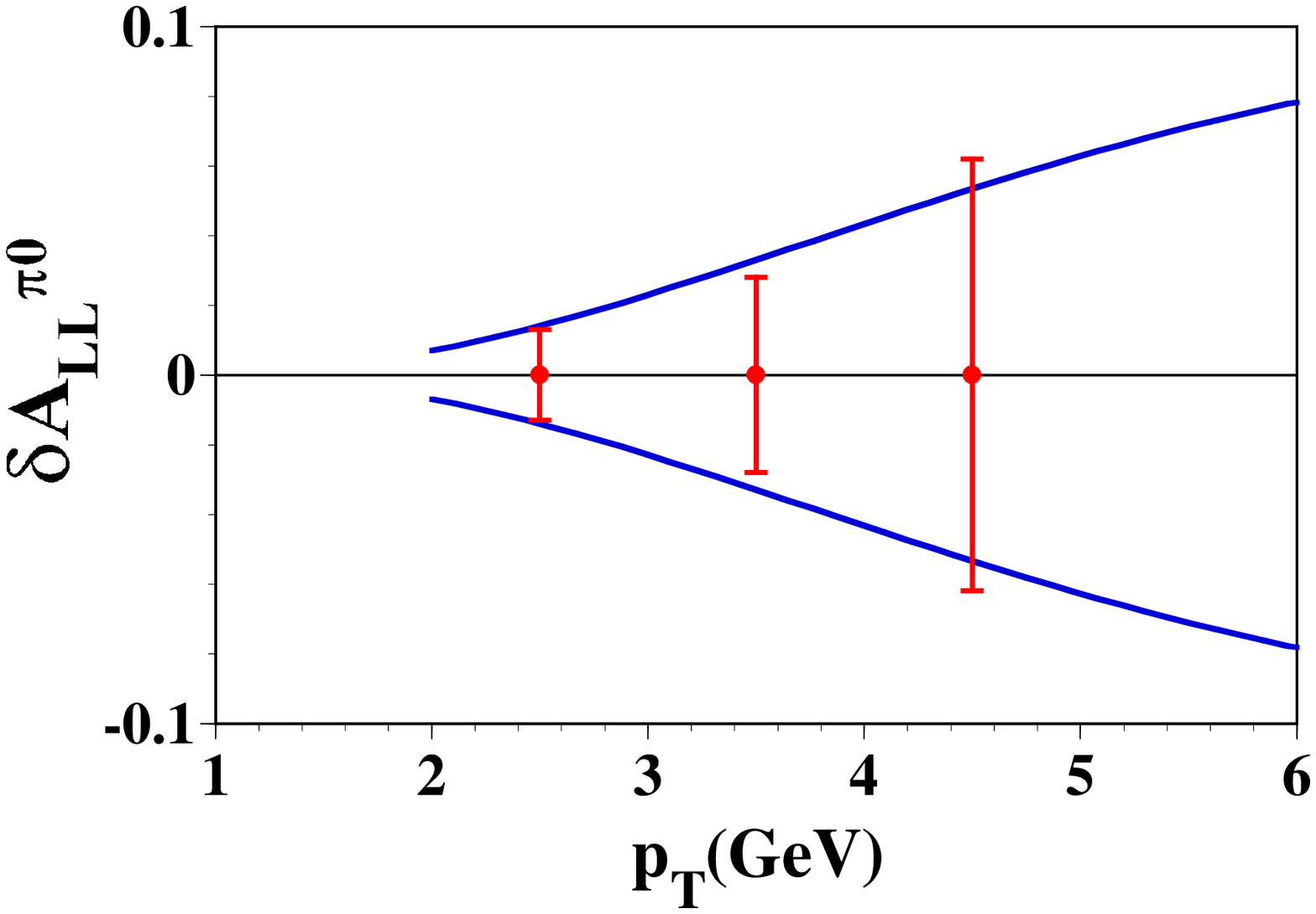} 
\caption{\label{fig:uncertainty}
Comparison of the asymmetry uncertainty $\delta A_{LL}^{\pi^0}$
with the statistical errors for 
$\sqrt{s}=200$ GeV.
}
\end{figure}
In Fig. \ref{fig:uncertainty}, the asymmetry uncertainty is compared to 
the statistical errors of the experimental data by the PHENIX \cite{PHENIX03_pol}.
In this comparison, we exclude the data at $p_T=1.5$ GeV. 
This is because that the data might have contribution from soft physics, 
and it might not be explained as physics of a hard process. 
We have no idea whether such data can be included in the global analysis. 
From this figure, 
we find that the uncertainty almost corresponds to the experimental errors,
and is mainly composed of the uncertainty of $\Delta g(x)$. 
This fact indicates that 
the present $\pi^0$ data have the same constraint on $\Delta g(x)$ 
as the polarized DIS data.
At this stage, one cannot expect to reduce the $\Delta g(x)$ uncertainty 
even if these data are included into the global analysis.
However, the asymmetry uncertainty is very sensitive to the $\Delta g(x)$ uncertainty. 
The $\pi^0$ production has the potential to become a good probe 
for $\Delta g(x)$ by future precise data.

It should be noted that symmetric uncertainty is shown 
in order to compare with the statistical errors in Fig. \ref{fig:uncertainty}. 
The lower bound is however incorrect because a lower limit of the asymmetry 
is not taken into account. 
As mentioned in previous subsection, 
the asymmetry cannot exceed $-0.1$ \% at low $p_T$ where the gg process dominates. 
Although asymmetric uncertainty should be estimated, 
such uncertainty cannot be obtained by the Hessian method. 
We therefore need further investigation of 
the lower bound for the asymmetry uncertainty.

\section{\label{chargedAsym}Spin asymmetry for charged pion production}
\begin{figure}[t]
\centering
\includegraphics[scale=0.42,clip]{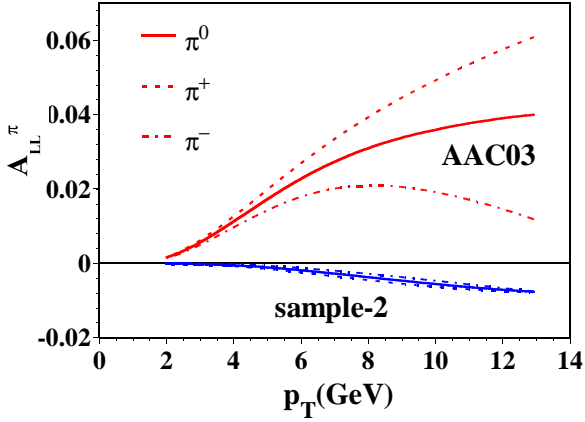}
\caption{
Asymmetries for neutral and charged pion productions 
with the AAC and sample-2 $\Delta g(x)$ sets.}
\label{fig-charged}
\end{figure}

We discuss the spin asymmetry for charged pion production, 
$\pi^+$ and $\pi^-$. 
Unpolarized and polarized cross sections can be similarly calculated 
by using the fragmentation functions decaying into charged pion $D^{\pi^\pm}$ 
in Eq. (\ref{eq:xsec}). 
We show asymmetries with the AAC $\Delta g(x)$ 
and sample-2 $\Delta g(x)$ in Fig. \ref{fig-charged}. 
In the asymmetries for the AAC $\Delta g(x)$, 
one can see differences among them in large $p_T$ region 
where the $qg$ process is dominant. 
The polarized cross sections of $qg$ process for $\pi^+$ and $\pi^-$ production
are written by
\begin{eqnarray}
    \Delta\sigma_{qg}^{\pi^\pm}&=&
        \Delta g \, \conv 
        \left(\sum_{i} \Delta f_i \, \conv \, D_{i}^{\pi^\pm} \right) 
        \conv \, \Delta \hat{\sigma}^{qg\to qg} \nonumber \\
     &+&\Delta g \, \conv 
        \left(\sum_{i} \Delta f_i \right) \conv \, D_{g}^{\pi^\pm} 
        \conv \, \Delta \hat{\sigma}^{qg\to qg} ~.
    \label{charged}
\end{eqnarray}
where the symbol $\conv$ denotes convolution integral in Eq. (\ref{eq:xsec}).
$i$ indicates the quark flavor, 
and is taken as $i=u, d, s, \bar{u}$, $\bar{d}$, and $\bar{s}$.
Actual calculation includes permutated terms of $x_a$ and $x_b$.
There are following relations among the fragmentation functions for charged pion:
\begin{equation}
\begin{array}{ll}
  D_u^{\pi^+}>D_u^{\pi^-},         & D_d^{\pi^+}<D_d^{\pi^-} , \nonumber \\
  D_q^{\pi^+}=D_{\bar{q}}^{\pi^-}, & D_g^{\pi^+}=D_g^{\pi^-} ,
\end{array}
\end{equation}
and the fragmentation functions for neutral pion are defined by 
\begin{equation}
  D_i^{\pi^0}=(D_i^{\pi^+}+D_i^{\pi^-})/2~.
\end{equation}
For $\pi^+$ production, the contribution associated with $\Delta u$ distribution 
is enhanced by the fragmentation function $D_u^{\pi^+}$. 
Increasing asymmetry for $\pi^+$ production is caused by positive contribution 
from $\Delta u$ distribution, 
whereas decreasing asymmetry  for $\pi^-$ production comes from 
negative $\Delta d$ distribution.  

On the other hand, 
the asymmetries for the sample-2 $\Delta g(x)$ are almost the same. 
The differences among them depends on the magnitude of $\Delta g(x)$, 
since the asymmetry is proportional to $\Delta g(x)$ 
as written in Eq. (\ref{charged}). 
If the absolute value of $\Delta g(x)$ is small, 
there are not significant differences among 
the asymmetries for $\pi^0$, $\pi^+$, and $\pi^-$ productions. 

In order to determine $\Delta g(x)$ with its sign 
by using charged pion production, 
let us propose an interesting observable which is defined by 
\begin{equation}
    A_{LL}^{\pi^{+}-\pi^{-}}=
        \frac{\Delta\sigma^{\pi^+ - \pi^-}}
             {\sigma^{\pi^+ - \pi^-}}
        \equiv\frac{\Delta\sigma^{\pi^+}-\Delta\sigma^{\pi^-}}
                   {\sigma^{\pi^+}-\sigma^{\pi^-}} ~. 
    \label{asym_pipm}
\end{equation}
The behavior of the asymmetry is sensitive to the sign of $\Delta g(x)$ 
because the contribution of the $gg$ processes are eliminated 
and one of the $qg$ process becomes dominant in the whole $p_T$ region. 
The polarized cross section for $gg \to gg$ process is given by 
\begin{equation}
    \Delta\sigma_{gg}^{\pi^\pm}= 
        \Delta g \, \conv \, \Delta g \, \conv \, D_{g}^{\pi^\pm} \,
        \conv \, \Delta \hat{\sigma}^{gg \to gg} .
\end{equation}
This contribution is cancelled out due to $D_g^{\pi^+}=D_g^{\pi^-}$. 
For the same reason, $gg\ra q\bar{q}$ process does not also contribute 
by summing fragmentation functions for flavors: 
$\sum_i D_i^{\pi^+}=\sum_i D_i^{\pi^-}$. 
As the similar case, the contributions of $q\bar{q}\ra gg$, 
$q\bar{q}\ra q'\bar{q}'$ processes are also vanished. 
The unpolarized cross section can be similarly calculated 
with unpolarized PDF's and partonic cross sections.

The asymmetry can be obtained by the difference of $qg$ process.
The second term of Eq. (\ref{charged}) is cancelled out 
for the same reason of $gg \to gg$ process. 
And then, the asymmetry is consequently given by 
\begin{equation}
A_{LL}^{\pi^{+}-\pi^{-}} \simeq
   \frac{\Delta g \,\conv
   \left(\Delta u_{v} - \Delta d_{v}\right) \conv
   \left(D_1^{\pi}-D_2^{\pi}\right) \conv \, \Delta \hat{\sigma}^{qg \to qg}}
   {g \,\conv \left(u_{v} - d_{v}\right) \conv
   \left(D_1^{\pi}-D_2^{\pi}\right) \conv \, \hat{\sigma}^{qg \to qg}} ~, 
\label{sigma_pipm}
\end{equation}
where $\Delta f_v(=\Delta f-\Delta \bar{f})$ is a polarized valence quark distribution.
The following relations among the fragmentation functions are assumed 
by the isospin symmetry, 
\begin{equation}
\left\{
    \begin{array}{ll}
    D_{u}^{\pi^+}=D_{\bar{d}}^{\pi^+}=D_{\bar{u}}^{\pi^-}=D_{d}^{\pi^-}
    &\equiv D_1^{\pi} \\ 
    D_{u}^{\pi^-}=D_{\bar{d}}^{\pi^-}=D_{\bar{u}}^{\pi^+}=D_{d}^{\pi^+}
    &\equiv D_2^{\pi}
    \end{array}
\right.
\end{equation}
This relation is used in parametrization of the fragmentation functions 
\cite{Kretzer00}.

In the asymmetry in Eq. (\ref{sigma_pipm}), 
ambiguity of fragmentation function $D_g^{\pi^\pm}$ is removed 
by the cancellation of the convolution part. 
Another ambiguity from the fragmentation functions can be also cancelled 
between numerator and denominator. 
In addition, $\Delta u_{v} - \Delta d_{v}$ is determined well, 
since its first moment is constrained 
by neutron and hyperon beta decay constants \cite{AAC03,BB,LSS,GRSV}.
Of course, unpolarized PDF's are precisely determined
in comparison with the polarized PDF's.
This asymmetry can be defined by well known distributions without $\Delta g$;
therefore, we can effectively extract information about $\Delta g(x)$ 
including its sign.

\begin{figure}[t]
\centering
\includegraphics[scale=0.42,clip]{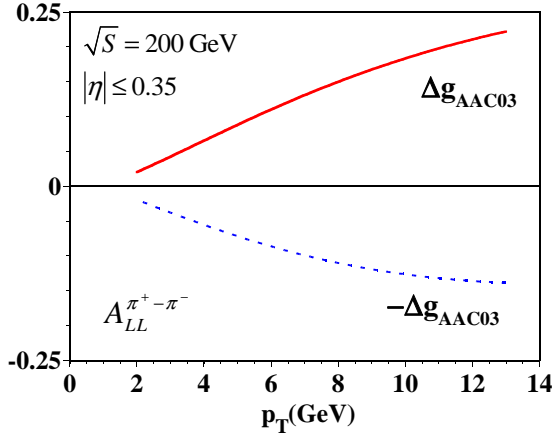}
\caption{
Asymmetries for the difference of charged pion production 
with $\Delta g(x)$ and $-\Delta g(x)$.}
\label{fig-asym_pipm}
\end{figure}
Figure. \ref{fig-asym_pipm} shows the asymmetry defined by Eq. (\ref{asym_pipm}). 
Solid and Dotted curves are asymmetries with AAC $\Delta g(x)$ 
and $-\Delta g(x)$. 
We find large asymmetries in both cases. 
In particular, the asymmetry with $-\Delta g(x)$ is negative and 
the absolute value is large in comparison with single pion production. 
Since the asymmetry is dominated by $qg$ process in the whole $p_T$ region, 
the difference of the sign of $\Delta g(x)$ is markedly reflected in the asymmetry.

We mention the contribution of $qq$ process to the asymmetry. 
In the region $8<p_T<13$ GeV, 
the $qq$ contribution accounts for 10-15\% of the polarized part 
($\Delta\sigma^{\pi^+}-\Delta\sigma^{\pi^-}$), 
and 27-56\% of the unpolarized part ($\sigma^{\pi^+}-\sigma^{\pi^-}$) 
of the asymmetry in Eq. (\ref{asym_pipm}). 
These contributions are not negligible. 
In particular, effect of the $qq$ contribution in the polarized part appears 
as the difference between the absolute values of asymmetries. 
Contributions of all sub-processes are taken into account, 
however the $q\bar{q}^{(\prime)}\ra q\bar{q}^{(\prime)}$ 
and $qq'\ra qq'$ processes are negligible. 
The difference is due to the positive contribution of $qq$ process. 
The asymmetry with $-\Delta g(x)$ is therefore suppressed. 

Next, we evaluate the experimental sensitivity of this spin asymmetry. 
We compare the statistical error of the asymmetry $\delta A_{LL}^{\pi^{+}-\pi^{-}}$ 
with one of $\pi^0$ production $\delta A_{LL}^{\pi^0}$. 
An expected statistical error $\delta A_{LL}^{\pi^{+}-\pi^{-}}$ is given by 
\begin{equation}
    \delta A_{LL}^{\pi^{+}-\pi^{-}}=\frac{1}{P^2}
    \frac{1}{\sqrt{N^{\pi^+}}}\frac{\sqrt{1+\alpha}}{1-\alpha} , 
\end{equation}
where $P$ is the beam polarization. 
$\alpha$ is the ratio of the number of event for $\pi^-$ and $\pi^+$: 
$\alpha=N^{\pi^-}/N^{\pi^+}$.
$N^{\pi^\pm}$ are obtained by the integrated luminosity ${\cal L}$ 
and the unpolarized total cross section $\sigma^{\pi^\pm}$: 
$N^{\pi^\pm}={\cal L}\sigma^{\pi^\pm}$. 
The ratio of these statistical errors can be obtained by 
\begin{equation}
  R_{sta}\equiv
  \frac{\delta A_{LL}^{\pi^{+}-\pi^{-}}}{\delta A_{LL}^{\pi^0}}
  =\frac{1}{\sqrt{2}}\frac{1+\alpha}{1-\alpha} ~. 
\end{equation}
The parameter $\alpha$ has energy dependence, and decreases with $p_T$ increases. 

Table. \ref{table1} represents the value of these parameters. 
$R_{asym}\equiv A_{LL}^{\pi^{+}-\pi^{-}}/A_{LL}^{\pi^0}$ is the ratio of asymmetries. 
The $\pi^+ -\pi^-$ asymmetry is about 5 times larger than the $\pi^0$ asymmetry. 
In the region $p_T<11$ GeV, the statistical error becomes larger than 
the rate of the asymmetry. 
The constraint power of experimental data are weaker than 
that of the $\pi^0$ asymmetry below the region. 
However, the value of $R_{asym}$ becomes larger than $R_{sta}$ above the region. 
The asymmetry would have the same impact on $\Delta g(x)$ as the $\pi^0$ production. 
Although more luminosity is needed in comparison with $\pi^0$ production, 
it is useful to determine effectively the behavior of $\Delta g(x)$ with the sign. 

\begin{table}[t!]
\centering
\caption{The value of parameters $\alpha$, $R_{sta}$, and $R_{asym}$.}
\label{table1}
\begin{tabular*}{8cm}{c|@{\extracolsep{\fill}}ccccc}
\hline
\hline
  \makebox[2.0cm]{$p_T$ (GeV)} & 9 &  10  &  11  &  12  &  13~~~~  \\ \hline
  $\alpha$    &  0.82   &  0.80  &  0.78  &  0.76  &  0.74~~~~  \\
  $R_{sta}$   &  7.2   &  6.4  &  5.7  &  5.2  &  4.7~~~~  \\
  $R_{asym}$  &  5.0   &  5.1  &  5.3  &  5.5  &  5.6~~~~  \\ 
\hline
\hline
\end{tabular*}
\end{table}

\section{\label{Summary} Summary}
In summary, we have investigated the correlation between 
the behavior of the spin asymmetry for pion production 
and the functional form of $\Delta g(x)$. 
The experimental data by the PHENIX indicates 
the negative asymmetry at low $p_T$, 
and motivate us to modify the functional form of $\Delta g(x)$ drastically.
In order to obtain negative asymmetry, 
the functional form of $\Delta g(x)$ requires some restraints.
By modifying $\Delta g(x)$, 
the slight negative asymmetry can be obtained at low $p_T$.
Moreover, we have indicated the existence of the negative polarization of $\Delta g(x)$ 
which keeps the asymmetry to be negative in the whole $p_T$ regions. 
The large negative asymmetry is inconsistent with the theoretical predictions 
by using $\Delta g(x)$ from polarized DIS data. 
However, experimental uncertainties are large at present.
It is premature to conclude that the pQCD framework is not applicable to 
$\pi^0$ production in polarized $pp$ collisions. 

Uncertainty of the $\pi^0$ asymmetry coming from the polarized PDF's with DIS data 
is correspond to the current statistical errors by the PHENIX. 
These data have the same constraint power on $\Delta g(x)$ as present DIS data. 
The future measurements will provide useful information for clarifying 
the gluon spin content.

Furthermore, we have proposed the spin asymmetry 
defined by the difference of cross sections for $\pi^+$ and $\pi^-$ production. 
We have discussed an impact of the asymmetry on determination of $\Delta g(x)$.
In the asymmetry $A_{LL}^{\pi^+-\pi^-}$, the $gg$ processes are cancelled out, 
and $qg$ process becomes dominant. 
Ambiguity of the fragmentation functions can be reduced.
The behavior of the asymmetry is sensitive to the sign of $\Delta g(x)$. 
One can obtain new probe for $\Delta g(x)$ in pion production at RHIC.

\def\Journal#1#2#3#4{{#1} {\bf #2}, #3 (#4)}
\def\NIM{Nucl. Instrum. Methods}
\def\NIMA{Nucl. Instrum. Methods A}
\def\NPB{Nucl. Phys. B}
\def\PLB{Phys. Lett. B}
\def\PRL{Phys. Rev. Lett.}
\def\PRD{Phys. Rev. D}
\def\ZPC{Z. Phys. C}
\def\EPJ{Eur. Phys. J. C}
\def\PR{Phys. Rept.}
\def\IJM{Int. J. Mod. Phys. A}

\end{document}